# Ambient Betatron Motion and Its Excitation by "Ghost Lines" in Tevatron[1]

**Vladimir Shiltsev**[a][*] **, Giulio Stancari**[a] **and Alexander Valishev**[a]

   [a] *Fermi National Accelerator Laboratory,*
   *PO Box 500, Batavia, IL, 60510, USA*
   *E-mail:* shiltsev@fnal.gov

ABSTRACT: Transverse betatron motion of the Tevatron proton beam is measured and analysed. It is shown that the motion is coherent and excited by external sources of unknown origins. Observations of the time varying "ghost lines" in the betatron spectra are reported.

KEYWORDS: Instrumentation for particle accelerators and storage rings - high energy (linear accelerators, synchrotrons)

---

[1] Work supported by Research Alliance, LLC under Contract No. DE-AC02-07CH11359 with the US Department of Energy.

[*] Corresponding author

# Contents

# 1. Introduction

Motion of the accelerator components, most notably, quadrupole magnets, results in the beam orbit movements and can lead to a significant deterioration of the collider performance in the Tevatron proton-antiproton collider. The mechanism depends on the frequency. At very high frequencies comparable or higher than the betatron frequency $f_0$ $(1-v) \approx 19.7$ kHz fluctuations of the magnetic fields $\delta B(t)$, e.g. due to quadrupole magnet displacements $x(t)$, produce transverse kicks $\delta\theta(t) = \delta B(t)el/Pc = x(t)/F$, where $l$ is length of the element, $F$ is the focusing length, that leads to the rms normalised beam emittance growth at the rate of [1]:

$$\frac{d\varepsilon_x}{dt} = \gamma \frac{f_0^2}{F^2} \sum_{k=1}^{N_q} \sum_{n=-\infty}^{\infty} \beta_k S_x(f_0(v-n)) \qquad (1)$$

where $f_0$ is the revolution frequency of 47.7 kHz in the Tevatron, $\gamma \approx 1044$ is the relativistic factor for the 980 GeV beams, $v$ is the tune, $S_x(f)$ is the power spectral density of the quadrupole motion $x$, $N_q$ is a total number of quadrupole focusing magnets, and $\beta_k$ is the beta function at the location of the element.

At much lower frequencies, $f \ll f_0$, the kicks lead to distortion of the closed orbit of the beams:

$$X_{COD}(s) = \frac{\sqrt{\beta(s)}}{2\sin(\pi v)} \sum_{k=1}^{N_q} \sqrt{\beta_k(s)}\theta_k \cos(\varphi(s) - \varphi_k + \pi) \qquad (2)$$

where $s$ is location along the ring, and $\varphi(s), \varphi_k$ are betatron phases at the locations of the observation point and at the source of the $k$-th magnet. At ultra-low frequencies, hours to years, the uncorrelated quadrupole magnet displacements are often governed by the *"ATL law"* [2,3] according to which the mean square of relative displacement $dX^2$ of the points separated by distance $L$ grows with the time interval between measurements $T$ as:

$$< dY^2 > = A\,T\,L \qquad (3)$$

where $A$ is a site dependent constant of the order of $10^{-5\pm1}$ μm$^2$/(s·m), and brackets <...> indicate averaging over many points of observations distanced by $L$ and over all time intervals equal to $T$. Such a wandering of the accelerator elements takes place in all directions. Corresponding average closed orbit distortion over the ring with circumference $C$ is equal to [4]:

$$< X_{COD}^2(s) > = \frac{\beta(\beta_F + \beta_D)}{8F^2\sin^2(\pi\nu)}ATC = \kappa ATC \quad (4)$$

where FODO focusing lattice structure is assumed, $\beta_F$, $\beta_D$ are beta functions at the focusing and defocusing lenses, and numerical coefficient $\kappa \approx 3$ and $A = (2-5)\times 10^{-6}$ μm$^2$/(s·m) [5] for the Tevatron. The orbit drifts by a fraction of a mm from empirically found "good" positions usually results in reduced lifetime of antiprotons and protons and untolerable losses at several tight aperture locations. Automated orbit control system is in place [6] while regular realignment of the magnets – usually during annual shutdown periods – helps to keep the orbit corrector currents within the limits.

## 2. Betatron Oscillations of Beam Centroid

### 2.1 Measurement technique

Several instruments are being used to detect orbit motion in the Tevatron. The most challenging is direct measurement of the minuscule betatron oscillations. Several special instruments were built for the purpose of detecting natural (ambient) beam oscillations and, therefore, determination of the betatron frequencies without additional excitations. Various techniques were employed, including 3D-BBQ (Direct Diode Detector Baseband tune measurement system (3D-BBQ) [7] and the Digital Tune Monitor (DTM) which uses 16 bits 100 MHz ADC's for measuring the tunes on a bunch-by-bunch basis [8]. Below we describe yet another, very high precision system that allows direct measurement of the betatron oscillations.

The system for the detection of the transverse coherent motion (Fig.1) is based on the signal from a single vertical beam position monitor (BPM) located near the CDF interaction point, in a region where the vertical beta-function at is $\beta_y = 900$ m. The BPM is a pair of 50 Ohm 18 cm long striplines pickups , each subtending 110 degrees of arc, with a circular aperture of 70 mm diameter and with two plate outputs ($A$ and $B$) for each of the two counter-propagating beams. The difference of the signals ($A$-$B$) is proportional to the beam displacement at the location of the BPM. The proton outputs are split: half of the signal is sent to the Tevatron BPM readout and orbit stabilization circuits; the other half is used by the present system. Antiproton signals are about a factor three weaker and are usually not used for orbit feedback, so the splitter is not necessary and the full signal can be analyzed. Switching between proton and antiproton signals presently requires physically swapping cables. In the Tevatron, protons and antiprotons share a common vacuum pipe. Outside of the interaction regions, their orbits wrap around each other in a helical arrangement. Therefore, bunch centroids can be several millimeters away from the BPM's electrical axis. Typically, the peak-to-peak amplitude of the proton signal is 10 V on one plate and 5 V on the other, whereas the signal of interest is of the order of a few millivolts. For this reason, it is necessary to equalize the *A* and *B* signals to take advantage of the full dynamic range of the digitizer. Equalization also reduces false transverse signals due to trigger jitter.

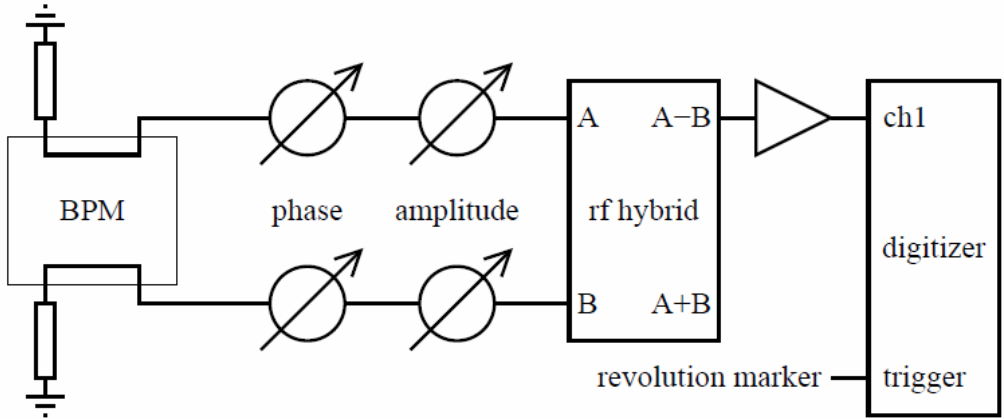

Fig.1: Schematic diagram of the high precision beam position measurement system

The phase and attenuation of each signal is manually adjusted by minimizing the *A-B* output of the rf hybrid circuit. If necessary, fine-tuning is done by displacing the beam with a small orbit bump. Orbits at collisions are stable within a mm over a time scale of weeks, and this manual adjustment does not need to be repeated often. To automate the task in the case of changing orbits and intensities (e.g., for observations at the top energy of 980 GeV between the low-beta squeeze and initiating collisions, or for observing both proton and antiproton bunches), a circuit board is being designed with self-calibrating gains and offsets. As the results, subtraction of the two signals by an RF hybrid provides about 44 dB of common mode suppression. The difference signal from the hybrid is amplified by 23 dB and sent to the digitizer. A single-channel, 1-V full range, 10-bit digitizer (Agilent Acqiris series) with time-interleaved ADCs is used. It can sample at 8 GS/s, so with sample period of 125 ps it corresponds to 150 slices for each 19-ns rf bucket, and store a maximum of 1024 MS or 125,000 segments. The 47.7-kHz Tevatron revolution marker is used as trigger. At 8 GS/s sampling rate, one can record waveforms of 1 bunch for 62,500 turns, 2 bunches for 52,707 turns, or 12 bunches for 12,382 turns, depending on the measurement of interest. A *C++* program running on the front-end computer controls the digitizer settings, including its delay with respect to the Tevatron revolution marker. Data is written in binary format. The output contains the raw ADC data together with the trigger time stamps and the delay of the first sample with respect to the trigger. Timing information has an accuracy of about 15 ps, and it is extremely important for the synchronization of samples from different turns. This system has also been used for studies of coherent beam-beam interaction modes [9].

## 2.2 Measured Betatron Motion

Fig.2 shows 21,400 turn (0.44 s) record of the vertical beam position at the VB11 BPM location.

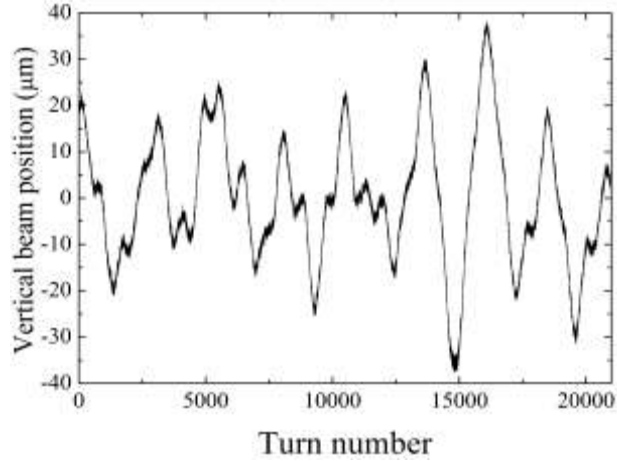

Fig.2: Vertical position of the proton bunch #11 at the beginning of HEP store #6214 (October 2008), measured at the VB11 location with $\beta_y$=900 m.

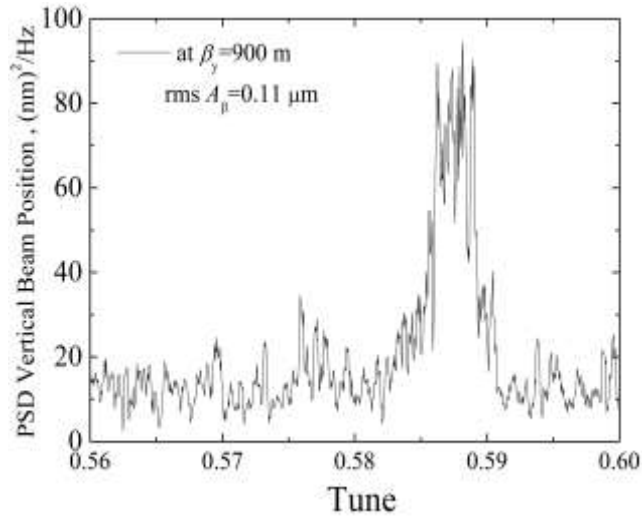

Fig.3: Power spectral density of the vertical betatron oscillations (FFT of the data presented in Fig.2).

The Fast Fourier Transform (FFT) of the data reveal significant excess of the signal at the betatron tune line over the noise as shown in Fig.3. To calculate properly the rms amplitude of the betatron oscillations one has to a) filter all harmonics except at the tune range 0.4-0.5 (by user of a FFT filter); b) make the FFT of the remaining signal; c) determine noise level (pedestal); d) subtract it from the signal at the betatron line; f) determine the signal level. At the end, the rms amplitude of the betatron oscillations is found to be about $A_\beta$ =110 nm. Note that the amplitude significantly varies in stores and from store to store (see next section) and often is 3-8 times smaller, that yields some 5-25 nm range of typical rms betatron motion amplitudes at the average beta function location with $\beta_y \approx$ 50 m. Even at that level, the detected rms amplitude is significantly higher than thermal motion (Schottky noise) of the proton bunch centroid, which can be estimated as $\delta \approx \sigma/N_p^{1/2} \approx$ 1nm, where $\sigma \approx 0.6$mm is the rms beam size and $N_p \approx 3 \times 10^{11}$ particles per bunch. Therefore, the motion is due to high frequency external forces.

Spectrum of the vertical orbit motion at frequencies 2-1000 Hz is shown in Fig. 4. It scales approximately as $\propto 1/f^3$ and is dominated by the low frequency beam motion. The strongest lines are the harmonics of 60 Hz main power. The 15 Hz and the 0.45 Hz components can be explained by the effects of the fast cycling Booster synchrotron and the Main Injector synchrotron on the power distribution systems at FNAL.

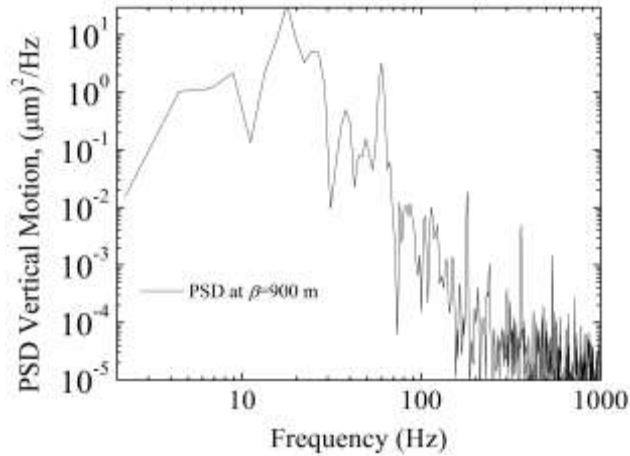

Fig.4: Low frequency power spectral density of the vertical orbit oscillations.

## 2.3 "Ghost Lines"

As noted above, the beam oscillation spectrum varies a lot, depending on conditions. The fastest and most widely operationally used instrument to monitor them is 21MHz Schottky detector [10]. The 21.4 MHz Schottky system is used to measure the horizontal and vertical tunes of the proton beam (antiproton signal is attenuated by 20 dB) without the possibility of gating on individual bunches. It has high resolution of about 0.0001.

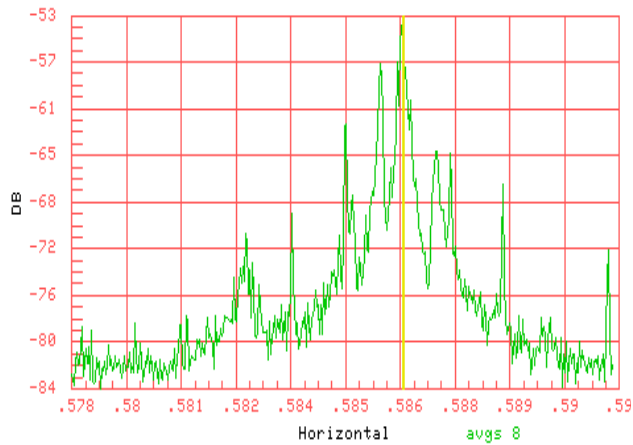

Fig.5: Horizontal 21 MHz Schottky spectrum of proton beam measured at 980 GeV measured 09/29/2002.

A typical Schottky spectrum from the 21.4 MHz pickups measured during collisions is shown in Fig.5. The spectrum contains many peaks, a set of lines around 0.586 are horizontal betatron tune lines (the vertical tune line was at 0.574 and is off the scale of the plot). Other remarkable peaks at 0.582, 0.584, 0.589 and 0.591 are what is called "ghost lines" corresponding to beam excitations of unknown origins. These lines were present in the Schottky spectra since early days of the Collider Run II and their most puzzling feature is that they are not stable but move around the frequency spectrum without any obvious reason. When they propagate through the vertical or horizontal oscillation frequencies, the power in the betatron motion goes up significantly.

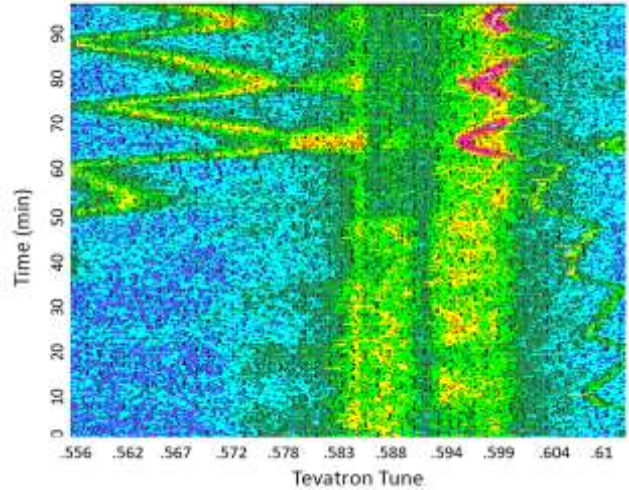

Fig.6: Horizontal 21 MHz Schottky spectrum of proton beam measured at 980 GeV (see text).

Fig.6 shows some 1.5 hour record of the Schottky spectrum during one of the colliding stores. Brighter lines are 0.596 and 0.585 indicate betatron bands, while several ghost lines were crossing them – coming both from below and from above. Note that their frequencies not only drift but also wiggle by as much as 0.005-0.02 with periods of about 15-20 minutes.

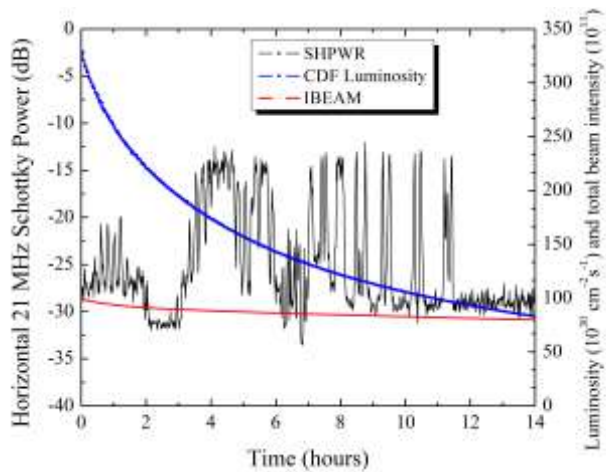

Fig.7: Variation of the horizontal 21 MHz Schottky power during 14-hours colliding run #8548 (March 9, 2011).

Fig.7 shows variation of the total Schottky power in the frequency band covering some 0.6 units of the tune space during a 14 hours long colliding store. One can see that the horizontal power varies by 18 dB , equivalent to the rms amplitude variation by a factor of 8.

## 3. Discussion and Conclusions

The direct measurement of the rms betatron oscillations amplitude estimates it to be about $A_{\beta}$=110 nm at $\beta_y \approx$900 m. Correspondingly, the amplitude at the average beta function location around the ring with $\beta_y \approx$50 m is about $A_{\beta} \times (50/900)^{1/2} \approx$25 nm. The betatron peak in the measured spectra is not always clearly seen, the power of the betatron oscillations varies significantly and we conclude that well known phenomenon of external excitation by so-called "ghost lines" is the reason – as these lines come and go without any obvious regularity. Our analysis of these "ghost lines" shows that a) the slowly wander across the frequency spectrum, with typical periods varying from 15-20 min to few hours; b) for the collider stores analyzed, the lines add about a factor of 2 to average (over colliding store duration) Schottky power in the betatron bands and, thus, contribute about 50% to the rms normalized emittance growth rate of 0.06 π mm mrad/hr [11]. The Tevatron "ghost lines" look very similar to infamous "humps" recently observed in the LHC. Those "humps" are unwanted oscillations seen repeatedly in the LHC beams (mostly in the vertical plane) and also believed to be caused by high frequency external excitations [12].

## Acknowledgments

We would like to thank A.Semenov, D.Still and J.Annala for technical support, assistance during beam studies, and many useful discussions on the subject of the "ghost lines".